\documentclass[aps,prl,superscriptaddress,draft,showpacs,intlimits,amsmath,amssymb,floats,floatfix,preprint]{revtex4}
\usepackage{bm}
\usepackage[final]{graphicx}
\usepackage{epsfig}
\begin{document}
\title{Pair breaking versus symmetry breaking: Origin of the Raman modes in superconducting cuprates}
\date{\today}
\author{N. Munnikes}
\affiliation{Walther Meissner Institute, Bavarian Academy of
Sciences, 85748 Garching, Germany}
\author{B. Muschler}
\affiliation{Walther Meissner Institute, Bavarian Academy of
Sciences, 85748 Garching, Germany}
\author{F. Venturini}
\altaffiliation[Permanent address: ]{Mettler-Toledo (Schweiz)
GmbH, 8606 Greifensee, Switzerland} \affiliation{Walther Meissner
Institute, Bavarian Academy of Sciences, 85748 Garching, Germany}
\author{L. Tassini}
\affiliation{Walther Meissner Institute, Bavarian Academy of
Sciences, 85748 Garching, Germany}
\author{W. Prestel}
\affiliation{Walther Meissner Institute, Bavarian Academy of
Sciences, 85748 Garching, Germany}
\author{Shimpei~Ono}
\affiliation{CRIEPI, Komae, Tokyo 201-8511, Japan}
\author{Yoichi Ando}
\affiliation{Institute of Scientific and Industrial Research,
Osaka University, Ibaraki, Osaka 567-0047, Japan}
\author{A. Damascelli}
\affiliation{Department of Physics {\rm \&} Astronomy, University
of British Columbia, Vancouver, BC V6T\,1Z4, Canada}
\author{H. Eisaki}
\affiliation{Nanoelectronic Research Institute, AIST, Tsukuba
305-8568, Japan}
\author{M. Greven}
\affiliation{Department of Applied Physics and Photon Science,
Stanford University, Stanford, CA 94305, USA}
\author{A. Erb}
\affiliation{Walther Meissner Institute, Bavarian Academy of
Sciences, 85748 Garching, Germany}
\author{R. Hackl}
\affiliation{Walther Meissner Institute, Bavarian Academy of
Sciences, 85748 Garching, Germany}
%\email{hackl@wmi.badw.de}

\begin{abstract}

We performed Raman experiments on superconducting ${\rm Bi_2 Sr_2
(Ca_{1-x} Y_x ) Cu_2 O_{8+\delta}}$ (Bi-2212) and ${\rm YBa_{2}
Cu_{3}O_{6+x}}$ (Y-123) single crystals. These results in
combination with earlier ones enable us to analyze systematically
the spectral features in the doping range $0.07 \le p \le 0.23$.
In $B_{2g}$ ($xy$) symmetry we find universal spectra and the
maximal gap energy  $\Delta_0$ to follow the superconducting
transition temperature $T_c$. The $B_{1g}$ ($x^2-y^2$) spectra in
Bi-2212 show an anomalous increase of the intensity towards
overdoping, indicating that the corresponding energy scale is
neither related to the pairing energy nor to the pseudogap, but
possibly stems from a symmetry breaking transition at the onset
point of superconductivity at $p_{\rm sc2} \simeq 0.27$.
\end{abstract}


\maketitle

Energy scales play an important role in solids whenever various
ground states are in close proximity. The copper-oxygen
superconductors %, in particular at low doping levels $p$,
are paradigmatic of competing order controlled by doping $p$. Yet,
the phases and relevant energies are still intensively debated
\cite{Damascelli:2003,LeTacon:2006,Tanaka:2006,Lee:2007,Gomes:2007,Hufner:2007,Alldredge:2008,Yoshida:2008}.
For example, in the underdoped range, $p \le 0.16$ holes per ${\rm
CuO_2}$ formula unit, the variation with $p$ of the
superconducting gap $\Delta_{\bf k}(p)$ is not settled. In some
experiments the maximum of the $d$-wave gap, $\Delta_0(p)$, is
found to increase or stay constant
\cite{Alldredge:2008,Yoshida:2008}. Other experiments indicate
$\Delta_0(p)$ to follow the superconducting transition temperature
$T_c$
\cite{Nemetschek:1997,Panagopoulos:1998,Deutscher:1999,Opel:2000,Sugai:2000,LeTacon:2006,Tanaka:2006,Lee:2007}.
A second energy $\Delta^{\ast}$ appears already at $T^{\ast} >
T_c$. $\Delta^{\ast}$ is the typical range over which spectral
weight is suppressed in the vicinity of $(\pi,0)$ and equivalent
points in the Brillouin zone (anti-node) and is usually referred
to as the pseudogap
\cite{Homes:1993,Loeser:1996,Timusk:1999,Hufner:2007}. $T^{\ast}$
and $T_c$ merge for $0.16 < p_{\rm m} < 0.20$ while there are
still two energy scales exhibiting different doping dependences
\cite{Nemetschek:1997,Panagopoulos:1998,Deutscher:1999,Timusk:1999,Damascelli:2003,Hufner:2007,Gomes:2007}.
There is general agreement that the one observed close to the Brillouin zone diagonal (node) %$(\pi/2,\pi/2)$
follows $T_c$. The anti-nodal one is approximately proportional to
$(1-p/p_0)$ with $0.16 < p_0 < 0.30$, similarly as
$\Delta^{\ast}(p)$. However, above $T_c$ there is no suppression
of spectral weight any more
\cite{Kendziora:1995,Nemetschek:1997,Panagopoulos:1998,Opel:2000,Sugai:2000,LeTacon:2006,Tanaka:2006,Gomes:2007,Hufner:2007,Lee:2007,Alldredge:2008,Yoshida:2008}
and coherence peaks are observed everywhere on the Fermi surface
by angle-resolved photoemission (ARPES) \cite{Damascelli:2003} and
in real space by scanning tunneling spectroscopy
\cite{Gomes:2007,Alldredge:2008}. The wide ranges of $p_{\rm m}$
and $p_0$ indicate systematic variations with both experiment and
material. In a recent Raman experiment the energy of the
anti-nodal pair-breaking peak was observed to decrease much faster
than $T_c$ upon applied pressure \cite{Goncharov:2003}.
Particularly the last result casts doubt on the prevailing
interpretation of the anti-nodal energy in terms of a direct
relationship to the pairing energy or the pseudogap.

In this paper, we present new electronic Raman scattering
experiments and put them into context with earlier results. We
systematically study the sample dependence and, as an additional
variable, the intensity of the superconductivity-induced features
for doping levels $0.07 \le p \le 0.23$. The results show that the
momentum dependence of the superconducting gap, $f({\bf
k})=\Delta_{\bf k}/\Delta_0$, hardly depends on doping for both
Y-123 and Bi-2212. At $p > 0.16$, the anti-nodal spectra of
Bi-2212 neither reflect the pseudogap nor the superconducting gap.
Rather, the doping dependence of both the intensity and the energy
of the superconductivity-induced modes suggests that they are
intimately related to the onset point of superconductivity at
$p_{sc2}=0.27$ on the very overdoped side of the phase diagram.

\begin{figure}[t]
  \centering
  \includegraphics [width=8.6cm]{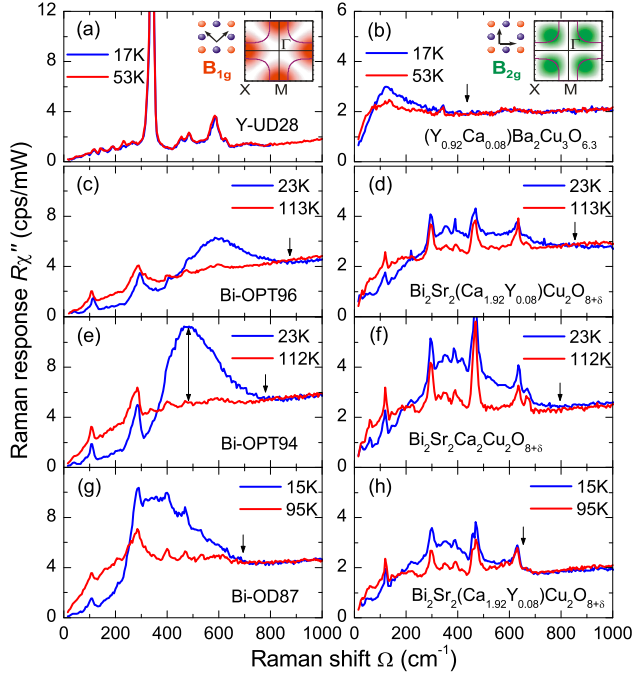}
  %\vspace{-0.8cm}
  \caption[]{Raman response $R\chi^{\prime\prime}(\Omega,T)$ (raw data) of
  ${\rm (Y_{0.92}Ca_{0.08})Ba_{2} Cu_{3}O_{6.3}}$ (Y-UD28) (a,b) and
  ${\rm Bi_2 Sr_2 ( Ca_{1-x} Y_x ) Cu_2O_{8+\delta}}$ (Bi-OPT94, Bi-OPT96, Bi-OD87) (c--h)
  in $B_{1g}$ and $B_{2g}$ symmetries as indicated. The corresponding light polarizations
  and sensitivities in the Brillouin zone are shown in
  the insets with copper and oxygen atoms displayed in red and
  blue, respectively. In (e) a double-headed arrow indicates the amplitude $A_{\rm sc}$ of
  the superconductivity-induced peak. Whenever applicable a
  down-pointing arrow gives the approximate position, where
  normal-state and superconducting spectra merge.

  } \label{fig:raw}
\end{figure}

Momentum-dependent electron dynamics such as gaps in
superconductors or collective modes can be studied by
Raman scattering.% since the photons excite intracell fluctuations
%of the charge density.
By appropriately adjusting the polarizations of the incoming and
outgoing photons different parts of the Brillouin zone can be
projected out independently \cite{Devereaux:2007}. In the
cuprates, $B_{1g}$ and $B_{2g}$ spectra emphasize anti-nodal and
nodal electrons, respectively, with the form factors shown in the
insets of Fig.~\ref{fig:raw} (a) and (b). In the superconducting
state the condensate is directly probed since the anomalous part
of the Green function is measured in addition to the normal one.
The spectra were measured with standard Raman equipment using the
Ar$^+$ line at 458~nm. The temperatures generally refer to the
illuminated spot and are typically between 5 and 10~K above those
of the holder.

In Fig.~\ref{fig:raw} we plot raw data of new measurements on
high-quality ${\rm (Y_{0.92}Ca_{0.08})Ba_{2} Cu_{3}O_{6.3}}$
(Y-UD28, $T_c=28$~K) (a,b), ${\rm Bi_2 Sr_2 Ca Cu_2O_{8+\delta}}$
(Bi-OPT94, $T_c=94$~K) (e,f), and ${\rm Bi_2 Sr_2 ( Ca_{0.92}
Y_{0.08} ) Cu_2O_{8+\delta}}$ (Bi-OPT96, $T_c=96$~K; Bi-OD87,
$T_c=87$~K) (c,d,g,h) single crystals. In spite of the almost
identical $T_{c}$s, the two optimally doped Bi-2212 samples
(Fig.~\ref{fig:raw} (c--f)) show substantial differences in the
$B_{1g}$ spectra. The peak energy $\Omega_{\rm peak}^{B1g}$ of
sample Bi-OPT96 is approximately 25\% higher than that of Bi-OPT94
while $T_c$ changes only by 2\%. The variation of the peak
position is accompanied by a change in the amplitude $A_{\rm sc}$,
i.e. the difference between the superconducting and the
normal-state spectra at the peak maximum, by a factor of 2.7.
These variations appear to be a result of subtle differences in
hole concentration and of quenched disorder \cite{Eisaki:2004}
leading to local strain fields. In the $B_{2g}$ spectra there are
only minor changes in shape, amplitude, and peak energy.

%%%%%%%%%%%%%%%%%%%%%%%%%%%%%%%%%%%%%%%%%%%%%%%%%%%%%%%%%%%%%%%%%%%
%%%%%%%%%%%%%%%%%%%%%%%%%%%%%%%%%%%%%%%%%%%%%%%%%%%%%%%%%%%%%%%%%%%

The overdoped sample Bi-OD87 [Fig.~\ref{fig:raw} (g,h)] was
prepared from Bi-OPT96 by oxygen annealing. Both peak frequencies
move downwards along with $T_c$ with a tendency of the $B_{1g}$
peak to move more rapidly than the $B_{2g}$ peak as observed
earlier in Bi-2212
\cite{Kendziora:1995,Opel:2000,Sugai:2000,Venturini:2002}, Y-123
\cite{Nemetschek:1997,Chen:1997,Nemetschek:1998,Opel:2000} and
$\rm HgBa_2CuO_{6+\delta}$ \cite{LeTacon:2006}.

On the underdoped side we studied Y-123 for its superior crystal
quality \cite{Erb:1996}. We find superconductivity to be
observable only in $B_{2g}$ symmetry. The peak energy is at
approximately one third of that observed at optimal doping and
follows $T_c$. The absence of superconductivity-induced peaks in
$B_{1g}$ symmetry appears to be a generic feature of underdoped
cuprates with $p \le 0.13$ (for a discussion see
ref.~\cite{Devereaux:2007}) which occurs in the same doping range
as the loss of coherence close to the anti-nodal points observed
in many experiments \cite{Venturini:2002b,Damascelli:2003}.

\begin{figure}[t]
  \centering
  \includegraphics [width=5.8cm]{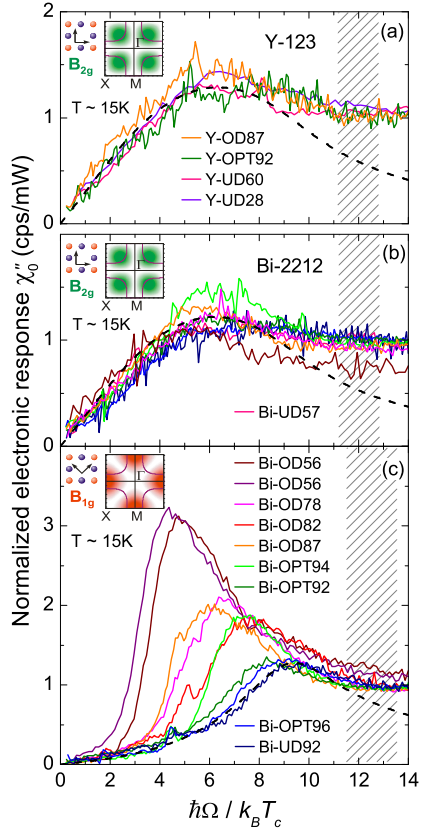}
  %\vspace{-0.8cm}
  \caption[]{Normalized electronic Raman response $\chi^{\prime\prime}_0(\Omega,p)$ of
  ${\rm (Y_{1-y}Ca_{y})Ba_{2} Cu_{3}O_{6+x}}$ in $B_{2g}$ symmetry (a) and
  ${\rm Bi_2 Sr_2 ( Ca_{1-x} Y_x ) Cu_2O_{8+\delta}}$ in
  $B_{2g}$ (b) and $B_{1g}$ symmetry (c). Spectra from samples other than those shown in
  Fig.~\ref{fig:raw} are taken from our published work
  \cite{Nemetschek:1997,Nemetschek:1998,Opel:2000,Venturini:2002}.
  For clarity, the phonons have been subtracted.
  The energy axes are normalized to the individual transition temperatures.
  All superconducting spectra merge with the normal-state response
  in the shaded range. %Note the universal linear
  %increase at low energy being a hallmark of a $d_{x^2-y^2}$
  %superconductor \cite{Devereaux:2007}.

  } \label{fig:scaling}
\end{figure}

It has been noticed earlier that the $B_{2g}$ peak energies in the
superconducting state follow $T_c$
\cite{Nemetschek:1997,Chen:1997,Deutscher:1999,Opel:2000,Sugai:2000,Venturini:2002,Devereaux:2007}.
Beyond that we demonstrate here that the entire $B_{2g}$ spectra
can be scaled by normalizing the energy axis of each sample to the
respective $T_c$ and the intensity to 1 at energies in the range
1000\,cm$^{-1}$. As shown in Fig.~\ref{fig:scaling}, the
superconducting $B_{2g}$ spectra collapse on universal curves for
both Y-123 and Bi-2212. The low-energy part of the normalized
spectra can be described quantitatively in terms of a
$d_{x^2-y^2}$ gap \cite{Devereaux:2007}. Naturally, the
description fails at higher energies since only the weak coupling
limit is considered which neglects the strong interactions
responsible for the large self energy of the electrons
\cite{Damascelli:2003} and, hence, the Raman spectra at high
energies \cite{Devereaux:2007}. With the gap represented by
  $  \Delta_{\bf k} = \Delta_0\left[{\cos{\bf k}_x-\cos{\bf
  k}_y}\right]/2$
we find agreement between theory and experiment up to the
pair-breaking peak (see Fig.~\ref{fig:scaling}). While the
$B_{2g}$ maximum $\Omega_{\rm peak}^{B2g}(p)$ itself scales as
$6~k_BT_c$ consistent with previous reports
\cite{Nemetschek:1997,Opel:2000,Sugai:2000,Venturini:2002,Devereaux:2007},
the gap maximum $\Delta_0$ from the $d$-wave fit in
Fig.~\ref{fig:scaling} follows $T_c$ as
  $2\Delta_0 = 9.0 \pm 0.5~ k_BT_c~$.
This demonstrates that both the gap ratio $\Delta_0/k_BT_c$ and
the momentum dependence $f({\bf k})$ remain unchanged in the
entire doping range studied and, hence, solves a long-standing
controversy.

%%%%%%%%%%%%%%%%%%%%%%%%%%%%%%%%%%%%%%%%%%%%%%%%%%%%%%%%%%%%%%%%%%%
%%%%%%%%%%%%%%%%%%%%%%%%%%%%%%%%%%%%%%%%%%%%%%%%%%%%%%%%%%%%%%%%%%%

The $B_{2g}$ results pose constraints on the interpretation of the
$B_{1g}$ spectra since the two symmetries are linked by the form
factors: a potential change of $f({\bf k})$ would inevitably leave
an imprint on  both $B_{1g}$ and $B_{2g}$ spectra. In turn, since
the $B_{2g}$ pair-breaking features remain invariant down to very
low doping the disappearance of the $B_{1g}$ gap structures for $p
< 0.15$ [Fig.~\ref{fig:raw}~(a)] cannot simply be traced back to
the gradual loss of quasiparticle coherence around the anti-nodal
parts of the Fermi surface
\cite{Gomes:2007,Lee:2007,Alldredge:2008} as proposed earlier
\cite{Chen:1997,LeTacon:2006}. In a sense, symmetry seems to
protect the nodal electrons. %But what happens at the anti-node?

In Fig.~\ref{fig:scaling}~(c) we plot electronic $B_{1g}$ spectra
of Bi-2212. As a general trend, the peaks move from 9.0 to
$4.5~k_BT_c$ for $p$ increasing from 0.15 to 0.23. While the
variation of the peaks is not monotonic, all normal and
superconducting spectra still merge in the same range of
approximately 12--14~$k_BT_c$ just as in $B_{2g}$ symmetry. % neither with
%doping nor with $T_c$.
%For example, the peaks of samples Bi-OPT94 and Bi-OPT96 are at 7.3
%and $9.0~k_BT_c$, respectively. Yet,

\begin{figure}[t]
  \centering
  \includegraphics [width=6cm]{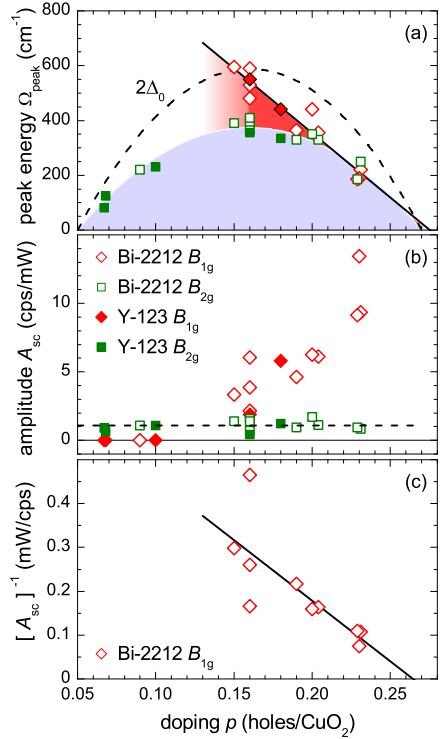}
  %\vspace{-0.8cm}
  \caption[]{Doping dependence of the superconductivity-induced features in Y-123 (full symbols) and
  Bi-2212 (open symbols).
  (a) Peak energies $\Omega_{\rm  peak}$. $\Omega_{\rm
  peak}^{B2g}$ (squares) is smaller than
  $2\Delta_0$ (dashes). The same holds true for $\Omega_{\rm
  peak}^{B1g}$ (diamonds) at $p > 0.16$. A linear fit (straight full line) represented by $\Omega_{\rm
  peak}^{B1g} = 1294(1-p/0.275)~{\rm cm}^{-1}$ extrapolates
  to the upper critical doping $p_{\rm sc2} \simeq
  0.27$ terminating the superconducting dome. (b) Amplitudes
  $A_{\rm sc}(p)$ in $B_{1g}$ and $B_{2g}$
  symmetries. The horizontal line at 1.03~cps/mW is the
  average of the amplitudes in $B_{2g}$ symmetry. (c) Inverse $B_{1g}$
  amplitudes $[A_{sc}(p)]^{-1}$ of Bi-2212. The linear fit
  extrapolates to zero at $p \simeq 0.26$ close to $p_{\rm sc2}$.

  } \label{fig:doping}
\end{figure}

In order to make connection to previous work
\cite{Kendziora:1995,Nemetschek:1997,Nemetschek:1998,Opel:2000,Sugai:2000,LeTacon:2006,Devereaux:2007}
we plot the peak energies $\Omega_{\rm peak}(p)$ for $B_{1g}$ and
$B_{2g}$ symmetry in Fig.~\ref{fig:doping}~(a). Also shown are
$2\Delta_0(p)$ and a linear fit to the $B_{1g}$ data. Clearly,
$\Omega_{\rm peak}^{B1g}(p)$ is unrelated to $2\Delta_0(p)$ while
the peak energies in $B_{2g}$ symmetry scale approximately as
$1.4\Delta_0(p)$ as expected from theory \cite{Devereaux:2007}.

As a new variable we analyze the amplitudes $A_{\rm sc}(p)$. In
Fig.~\ref{fig:doping}~(b) we compile results for $A_{\rm sc}(p)$
from the present study and our earlier results in Y-123
\cite{Nemetschek:1997,Nemetschek:1998,Opel:2000} and Bi-2212
\cite{Nemetschek:1997,Opel:2000,Venturini:2002} with all
amplitudes given in absolute units. The differences between Y-123
and Bi-2212 are small indicating little individual variation for
these two double-layer compounds and little influence of
resonantly enhanced scattering with excitation in the visible
spectral range \cite{Venturini:2002}. For $B_{2g}$ symmetry
$A_{\rm sc}(p)$ is practically doping independent with an average
close to 1~cps/mW. The variations of order $\pm 50\%$ between
individual samples not only reflect impurity effects
\cite{Devereaux:1995} but also variations of the overall cross
section which are not yet understood. Similar sample-dependent
changes are also observed in $B_{1g}$ symmetry. However, the large
basis of results allows us to derive two significant trends: (i)
below $p \simeq 0.13$, $A_{\rm sc}(p)$, i.e. any
superconductivity-induced spectral change, vanishes in $B_{1g}$
symmetry (cf. Fig.~\ref{fig:raw} (a)). This goes along with the
rapid decrease of the coherence peaks in tunneling
\cite{Gomes:2007} and in ARPES at the anti-nodal Fermi surface
crossing \cite{Damascelli:2003,Tanaka:2006,Lee:2007}. (ii) In
Bi-2212 $A_{\rm sc}(p)$ increases strongly for $p>0.18$ which has
not been appreciated yet. The two points from Y-123 follow the
same trend. If we plot $[A_{sc}(p)]^{-1}$ (Fig.~\ref{fig:doping}
(c)) we find a divergence point at $0.26 \pm 0.03$ close to
$p_{\rm sc2} = 0.27$ where superconductivity disappears (or
appears, depending on the point of view).

%%%%%%%%%%%%%%%%%%%%%%%%%%%%%%%%%%%%%%%%%%%%%%%%%%%%%%%%%%%%%%%%%%%
%%%%%%%%%%%%%%%%%%%%%%%%%%%%%%%%%%%%%%%%%%%%%%%%%%%%%%%%%%%%%%%%%%%

Given the universality of $f({\bf k})$ and $2\Delta_0/k_BT_c$, the
variation of $\Omega_{\rm peak}^{B1g}(p)/k_BT_c$ by a factor of
two, and the tendency of $A_{\rm sc}^{B1g}(p)$ to diverge, it is
hard to identify the $B_{1g}$ maximum with $\Delta_0$. What are
the alternatives?

An explanation in terms of an exciton-like bound state below
$2\Delta_0$ (Bardasis-Schrieffer mode) has been proposed recently
\cite{Chubukov:2006,Chubukov:2008}. At first glance, the energy
and intensity variations predicted on the basis of a
spin-fluctuation model are similar to those observed here with a
simultaneous increase of both amplitude and split-off below
$2\Delta_0$. However, the doping dependence of the $B_{1g}$ Raman
mode is just opposite to what one expects for the spin channel.
Similar arguments apply for a bound state induced by charge
ordering \cite{Zeyher:2004}. At present we are not aware of an
interaction with dramatically increasing coupling strength towards
high doping. Alternatively, band structure effects may play a
role. However, the quite complicated multi-sheeted Fermi surface
of Y-123 seems to have only little influence on the spectra in the
superconducting state.

Since these more traditional possibilities fail to provide a
qualitatively correct description of the experiments we explore a
scenario which rests on the unconventional evolution with doping
of the $B_{1g}$ intensity in Bi-2212. If individual variations
between the samples are neglected, $A_{\rm sc}^{B1g}(p)$ diverges
approximately as
  $A_{\rm sc}^{B1g}(p) \propto \left[ 1-{p}/{p_{\rm sc2}}
  \right]^{-1}$.
Although there is a substantial increase of anti-nodal coherence
in the ARPES single-particle spectra upon overdoping, such as in
the case of heavily overdoped Tl-2201 \cite{Plate:2005}, the
evolution of the $B_{1g}$ Raman response can hardly be explained
in this way. If this were the case, the $B_{1g}$ maximum would
just become sharper while conserving the integrated area. The
observed intensity increase along with the reduction of
$\Omega_{\rm peak}^{B1g}(p)$ [Fig.~\ref{fig:doping}~(a)] is
instead more compatible with the behavior of a Goldstone mode
appearing when a continuous symmetry is broken. In fact, we find
not only $A_{\rm sc}^{B1g}(p)$ to diverge at $p=0.26 \pm 0.03$ but
also $\Omega_{\rm peak}^{B1g}(p)$ to extrapolate linearly to zero
at $p=0.27 \pm 0.02$ as expected for a symmetry-breaking mode.

In this scenario the $B_{1g}$ spectrum is a superposition of the
weak coupling pair-breaking feature  and an additional mode with
$B_{1g}$ symmetry originating from a broken symmetry. The mode
depends on doping and, in close correspondence to the variation
with pressure of $\Omega_{\rm peak}^{B1g}/k_BT_c$
\cite{Goncharov:2003} on sample details \cite{Eisaki:2004}. The
microscopic origin remains open yet. A spin-density modulation
with ${\bf q}= (\pi,\pi)$ would not appear in $B_{1g}$ but,
rather, in $A_{1g}$ symmetry \cite{Venturini:2000}. A Pomeranchuk
instability of the Fermi surface \cite{Metzner:2003} or spin
and/or charge ordering fluctuations with $(0.2\pi,0)$
\cite{Wakimoto:2004,Caprara:2005} have the proper symmetry.

In conclusion, the doping independence of the normalized $B_{2g}$
pair-breaking spectra pins down the superconducting gap's momentum
dependence. The variations of energy and amplitude of the
superconductivity induced $B_{1g}$ spectra cannot originate from a
doping dependence of the gap, since there should also be an
influence on the $B_{2g}$ spectra. For $p > 0.16$ we are dealing
apparently with a mode of well defined $B_{1g}$ symmetry (typical
for a collective mode) rather than a projection of the gap as in
$B_{2g}$ symmetry. We speculate that at least in Bi-2212 the mode
indicates a broken continuous symmetry at the onset point of
superconductivity at $p_{\rm sc2 }\simeq 0.27$.

\textbf{Acknowledgements:} We thank Guichuan Yu for valuable
discussions and comments. We acknowledge support by the DFG under
grant numbers HA 2071/3 and ER 342/1 via Research Unit FOR538. The
crystal growth work at Osaka and Stanford was supported by KAKENHI
19674002 and by DOE under Contracts No. DE-FG03-99ER45773 and No.
DE-AC03-76SF00515, respectively.

\end{document}